\documentstyle[12pt]{article}
\begin{document}
\newcommand{\be}{\begin{equation}}
\newcommand{\ee}{\end{equation}}

\bigskip
\centerline {\bf Log-periodic Oscillations for Biased Diffusion on Random
Lattice}

\bigskip
\centerline {Dietrich Stauffer$^1$ and Didier Sornette$^2$}

\noindent
$^1$ Institute for Theoretical Physics, Cologne University, 50923 K\"oln,
Germany

\centerline {e-mail: stauffer@thp.uni-koeln.de}

\noindent
$^2$  Department of Earth and Space Science and Institute of Geophysics and
Planetary Physics, University of California, Los Angeles, California 90095, USA
and Laboratoire de Physique de la Mati\`ere Condens\'ee, CNRS UMR6622 and
Universit\'e de Nice-Sophia Antipolis, Facult\'e des Sciences, B.P. 71
06108 NICE Cedex 2, France

\centerline {e-mail: sornette@cyclop.ess.ucla.edu}

\bigskip
\bigskip
ABSTRACT:
Random walks with a fixed bias direction on randomly diluted cubic lattices
far above the percolation threshold exhibit log-periodic oscillations in the
effective exponent versus time. A scaling argument
accounts for the numerical results in the limit of
large biases and small dilution and shows the importance of the interplay
of these two
ingredients in the generation of the log-periodicity.
These results show that log-periodicity is the dominant effect compared to
previous
predictions of and reports on anomalous diffusion.

\bigskip \bigskip

Diffusion on percolating lattices has a long history [1-4]. Here the sites of
a large lattice are randomly initialized as being accessible, with probability
$p$, or forbidden, with probability $1-p$; and random walker diffuse on the
accessible sites only. In biased diffusion [5,6], with a probability $B$ the
random walk moves in one fixed direction, while with probability $1-B$ it
moves as usual to one randomly selected nearest neighbor; in both cases the
move is allowed only if the neighbor is accessible. This fixed bias corresponds
to an external electric field and is different from the bias with changing
direction depending on the flow through the random network [7].

Numerical searches for a possible phase transition as a function of the bias
$B$ were hampered by strong variations of the effective exponent $k$ in
$$<r^2> \propto t^k \eqno (1)$$
as a function of time $t$, for the mean square displacement $r$. In particular,
Seifert and Suessenbach [8] found smooth variations, if $k$ was plotted versus
the logarithm of time. Log-periodic oscillations were predicted before for
a special one-dimensional diffusion problem [9] while ref.8 investigated three
dimensions.

In the meantime, both computer technology and theoretical understanding have
advanced appreciably. Log-periodic oscillations were seen in many cases,
as reviewed recently [10], including the stock market crashes on Wall Street
in 1929 and 1987 [11] and in 1997 [12]. The physical mechanism for
log-periodicity obtained in [9] is the exponential dependence of the
trapping time on
the length of the traps that are multiples of the lattice mesh size.
The resulting intermittent random walk is thus punctuated by the successive
encounters with larger and larger clusters of trapping sites. For strong
bias $B$ in three dimensions,
we might expect a somewhat similar mechanism as discussed below. We were thus
motivated to repeat the work of ref.[8] for longer times, in search
for periodicities like $\sin(\log(t))$.

Each of the 512 processors of the Cray-T3E deals with about 3 million steps
per second, roughly the same speed as the supercomputer CDC Cyber 205 used in
ref.[8].
We could also simulate larger simple-cubic lattices ($301 \times 301 \times 301
$) even though we wasted a whole 32-bit word for each lattice site (instead of
one bit only) to simplify and accelerate the program (93 Fortran lines,
available from the first author). We averaged over about 500 lattices,
with 1000
walkers on each for short times and 3 for long times up to $4 \times 10^{10}$.

If the bias equals $B=1$, the motion is one-dimensional, and thus because
of ref.[9] the chances for log-periodic oscillations are best for high $B$.
Indeed, ref.[8] found clearer oscillations for $B = 0.99$ than for $B = 0.9$.
Fig.1 shows our results for $B=0.98$
giving very clearly three maxima in
$k \; vs. \; \log(t)$ near $t = 10^4, \; 2~10^6, \; 4~10^8$. Fig. 2 shows our
results for $B= 0.999$ giving two maxima  in
$k \; vs. \; \log(t)$ near $t = 10^5, \; 10^9$. Fig.3  shows our
results for $B= 0.95$ giving many more maxima but with much smaller
amplitude which
are thus harder to distinguish from noise.
As a function of the bias $0.9 < B < 0.999$ at fixed concentration $p=0.725$
we found slower oscillations for larger $B$, while for smaller $B$ they became
more rapid. For smaller $B$'s, the oscillations become masked by the random
fluctuations. We could probably attempt to measure them by a suitable averaging
procedure [10,13] and leave this for the future.

\vskip 0.5cm
Table 1\,: The table summarizes our estimates for the scaling factors
$\lambda$ defined as the ratio of the times of the successive maxima.

\begin{table*}[h]
\begin{center}
\begin{tabular}{|c|c|c|} \hline
B &  $\lambda$ & (1+5B)/(1-B)    \\ \hline
0.90 & $10 \pm 5$ & $55$\\ \hline
0.95 & $15 \pm 5$ & $110$ \\ \hline
0.98 & $200 \pm 100$ & $300$ \\ \hline
0.99 & $10^3$ & $6~10^2$  \\ \hline
0.999 & $10^4$ & $6~10^3$  \\ \hline
\end{tabular}
\end{center}
\end{table*}

We now present the scaling argument proposed to account for these
results, that should
be correct in the limit $B \to 1$ and $1-p \to 0$ but non zero. We expect
its domain
of validity to be much broader than these limits would indicate, in a way
similar to what we found for the scaling argument developed in [14] (see
also [10])  to
retrieve the exact
analytical results derived in [9]. The mechanism is best understood by
noticing that
the biased random walk along $x$ is the same as an unbiased random walk in
the presence of
a linearly decreasing potential
\be
V(x) = - b x~,
\label{xxccv}
\ee
where the slope $b$ is such that the ratio of
the forward over backward motion rate parallel to the bias is correctly
evaluated.
In the definition involving $B$, the forward rate is
${1-B \over 6} + B = {1+5B \over 6}$ (the unbiased contribution plus the
bias) and
the backward rate
is ${1-B \over 6}$. Equating their ratio to $\exp[\beta~b]$ yields $b$ in
(\ref{xxccv}),
where $\beta$ is the inverse temperature in the proposed analogy\,:
\be
b = {1 \over \beta} \log {1 + 5B \over 1-B}~~.
\label{rrriio}
\ee
$b \propto B$ for small $B$ and $b \propto \log {1 \over 1-B}$ for $B$
close to one.

As the random walker makes its way through the system, it will eventually
encounter a local well,
defined by a connected peninsula surrounded by empty sites (this can also be
called a dead end) such that it has to retrace its steps backward to
escape. Call $n$ the depth of this well. The typical time of residence of
the walker in this well (trapping time) is proportional to the Arrhenius factor
$\exp [\beta~b~n]$. This is the mechanism that converts a linear increase
(in $n$)
into an exponential increase in trapping times and thus produces log-periodic
oscillations. Indeed, at short times, it is more probable that the smallest
well
be first encountered. This is due to the fact that such trapping wells
occur with
small probability (see below), denoted $P_n$.
The smallest wells are of depth $n=1$ and lead to a trapping time
$\propto \exp [\beta~b~] = {1 + 5B \over 1-B}$. The next most severe
trapping wells
are of depth $n=2$ and lead to a typical trapping time
$\propto \exp [2 \beta~b~] = [{1 + 5B \over 1-B}]^2$. And so on. We see a
hierarchy
of time scales $[{1 + 5B \over 1-B}]^n$ resulting from the succession of worst
encountered trapping wells. Note that after it has encountered a well of
depth $n$, the walker
will in general encounter several wells of the same or smaller depths before
being trapped in a well of depth $n+1$. This does not destroy the argument 
because only the worst trapping well encountered until now dominates the 
waiting times and thus
the log-periodic modulations. This is completely similar to what was found
in the
one-dimensional case [9,10]. Note that we also predict that only a finite
number $n_{max}$ of
oscillations will be observed in a finite system of linear size $L$.
This simply stems from the finiteness of the deepest well $n_{max}$ in a finite system,
given by the condition $P_{n_{max}} ~L^d \sim 1$ (J. Machta and A. Aharony, 
priv.comm.). If
$P_n$ is exponentially decreasing with $n$, we see that $n_{max} \sim \log
L$, while it
is even smaller $n_{max} \sim [\log L]^{1 \over d-1}$ according to our
suggestion below.
This size effect has been verified qualitatively for
$B = 0.98$, with the same concentration $p = 0.725$ and sizes $L= 51, 21,
11$ and $5$.

Our theory teaches us that only the depth of the wells control the
leading behavior of the trapping time
scales. Their transverse sizes bring in only subleading power law
corrections to the
exponential Arrhenius dependence. This is the fundamental reason for the
existence and
robustness of the observed log-periodic oscillations, notwithstanding the many
random configurations that the trapping wells can take. Notice that, as in
all other
known cases [10], the log-periodic oscillations rely on the existence of an
ultraviolet
(or infrared) cut-off, here the mesh size.

We have reported in the last
column of table 1 the value of the scaling factor
\be
\lambda \equiv {t_{n+1} \over t_n} = {1 + 5B \over 1-B}~,
\label{eq2}
\ee
derived from (\ref{rrriio}). The agreement is quite good for the largest
bias $B \to 1$
for which the first smallest wells dominate completely the dynamics. For
smaller $B$'s,
corrections include the effect of all subleading time scales coming from all
well shapes of a given depth.
There is also a larger uncertainty in the determination of the scaling ratio
$\lambda$ from the figures as the log-periodic oscillations are weaker and
harder
to distinguish from statistical noise. The small ratios seen for the
smaller $B$'s
in the figures could also be higher harmonics.

This is all we need to account for the observed log-periodicity. Let us now
briefly comment on the distribution $P_n$ of trapping well sizes of depth $n$.
A trapping well of depth $n$ has a typical transverse size $n$ (isotropy).
It can be constructed in a way similar to that leading to the distribution of
finite cluster size above percolation. In this latter case, a finite
cluster $s$
is constructed from the infinite incipient percolation network by
introducing of the
order of $s^{d-1 \over d}$ empty perimeter sites. This leads to the stretched
exponential distribution $\exp [-a ~s^{1-1/d}]$ known to be exact [15].
In order to construct a trapping well of depth $n$, one needs also to introduce
of the order of $s^{1-1/d}$ empty perimeter sites. This would predict
the
distribution $\exp [-a' ~n^{d-1}]$, which decays faster than an exponential
in dimensions larger
than two. This is in contrast to [6] which used an exponential distribution
to predict
a transition
from drift to no drift (vanishing velocity) above a critical value of the
bias $B$.
Our results show that
this effect is very hard to detect numerically if true and that the main
effect of the trapping wells is by far in creating a hierarchy of time scales
$t_n$ given by (\ref{eq2}) leading to log-periodic oscillations
that completely dominates the dynamics.

To summarize,  damped oscillations, periodic in
the logarithm of the time, have been discovered for strongly biased
diffusion on disordered
three-dimensional lattices; these log-periodic oscillations stem from an
Arrhenius
exponentiation of a discrete spectrum of trapping well depths, leading to
a discrete hierarchy of time scales. Similar effects were
found for concentration $p=0.5$, while for bias $B=0.5$ no oscillations were
seen. Our result underlines the importance of accounting for intermittent
distortions
that will bias the determination of exponents whenever log-periodicity occurs.
We note that a similar difficulty for determining the fractal dimension of
fractal DLA clusters in the past has been attributed to the existence of
log-periodic modulation of the apparent exponent [16].

\bigskip
We are indebted to M. Barma and D. Dhar for pointing out an error
in a previous version and for useful correspondences.
We thank the Les Houches Spring School 1997 ``Scaling and Beyond'' [17] as
well as
L. Sch\"afer and his group at Essen University and J. Machta and A. Aharony
for crucial questions and
suggestions. The simulations were made at HLR J\"ulich in several thousand
processor hours.

\parindent 0pt
\bigskip
\bigskip
\bigskip
\newpage
[1] P. G. de Gennes, La Recherche 7, 916 (1976).

[2] C. D. Mitescu H. Ottavi and J. Roussenq, AIP Conference Proceedings 40,
 377 (1978)

[3] Y. Gefen, A. Aharony, S. Alexander; D. Ben-Avraham and S. Havlin, J. Phys.
A 15, L 691 (1992).

[4] S. Havlin and D. Ben Avraham, Adv. Phys. 36, 395 (1987)

[5] R.B.Pandey, Phys.Rev. B 30, 489 (1984);
H. B\"ottger and V. V. Bryskin, Phys. Stat. Sol.  (b) 113, 9 (1982).

[6] M. Barma and D. Dhar, J. Phys. C 16, 1451 (1983).

[7] A.Bunde, S. Havlin and H.E.Roman, Phys.Rev. A 42, 6274 (1990) and earlier
papers of this group cited there.

[8] E. Seifert and M. Suessenbach, J.Phys.A 17, L 703 (1994)

[9] J. Bernasconi and W.R. Schneider, J.Phys. A 15, L 729 (1982)

[10] D.Sornette, Physics Reports, in press
(http://xxx.lanl.gov/abs/cond-mat/9707012)

[11] D. Sornette, A. Johansen and J.-P. Bouchaud, J.Phys.I France 6, 167
(1996);
D. Sornette and A. Johansen, Physica A 245, 411 (1997)

[12] Prediction of the stock market turmoil at the end of october 1997,
 based on an unpublished extension of the theory,
 have been formally issued ex-ante on september 17, 1997,
to the French office for the  protection of
proprietary softwares and inventions under number registration 94781.
In addition,
a trading strategy has been devised
using put options in order to provide an experimental test of the
theory. A $400\%$ profit has been obtained in a two week period covering
the mini-crash of
october 31, 1997. The proof of this profit is available from a Merrill Lynch
client cash management account released in november 1997.
See also H. Dupuis, ``Un krach avant novembre'',
Tendances, 18. September 1997,
page 26, from the work of N. Vandewalle, A.Minguet, P.Boveroux, and M. Ausloos
using the same type of log-periodic signals.

[13] A. Johansen and D. Sornette, ``Canonical averaging'' for
log-periodicity, preprint

[14] H. Saleur, C.G. Sammis and D. Sornette, J. Geophys. Res.
101, 17661 (1996)

[15] H. Kunz and Souillard, B., J. Stat. Phys. 19, 77 (1978).

[16] D. Sornette, A. Johansen,  A. Arn\'eodo, J.-F. Muzy and H. Saleur,
Phys. Rev. Lett. 76, 251 (1996).

[17] B. Dubrulle, F. Graner and D. Sornette, eds.,
Scale invariance and beyond, Proceedings of Scale invariance and Beyond,
Springer, Heidelberg, 1997.

\bigskip
\bigskip
\bigskip

Fig.1: Effective exponent $k$ = d log $r^2$ / d log $t$ versus time $t$ for
$B=0.98$.
(combination of different runs with different statistics).

\bigskip
Fig.2: Same as Fig.1 for $B=0.999$.

\bigskip
Fig.3: Same as Fig.1 for $B=0.95$.

\end{document}